# Mach-Zehnder Bragg interferometer for a Bose-Einstein Condensate

**Yoshio Torii,\* Yoichi Suzuki, Mikio Kozuma and Takahiro Kuga**

*Institute of Physics, University of Tokyo, 3-8-1 Komaba, Meguro-ku, Tokyo 153-8902, Japan.*

**Lu Deng and E.W. Hagley**

*National Institute of Standards and Technology, Gaithersburg, MD 20899, USA.*

**Abstract:**

We construct a Mach-Zehnder interferometer using Bose-Einstein condensed rubidium atoms and optical Bragg diffraction. In contrast to interferometers based on normal diffraction, where only a small percentage of the atoms contribute to the signal, our Bragg diffraction interferometer uses all the condensate atoms. The condensate coherence properties and high phase-space density result in an interference pattern of nearly 100% contrast. In principle, the enclosed area of the interferometer may be arbitrarily large, making it an ideal tool that could be used in the detection of vortices, or possibly even gravitational waves.

**Paper:**

With the advent of Bose-Einstein condensation (BEC) of dilute atomic gases [1,2] comes a coherent matter wave, or *atom laser* [3-5], analogous to an optical laser. The high phase-space density and coherence properties of condensates make possible atom-optic experiments that have previously only been performed with optical lasers, such as four-wave mixing [6]. Among these, atom-laser (matter-wave) interferometers [7, 8] are of particular interest because of their potential high sensitivity. In order to be able to utilize atom lasers as one does optical lasers, atom-wave versions of optical elements, such as 50/50 beam splitters and mirrors, are needed. The recent demonstration of optical Bragg diffraction of a BEC by a moving, optical, standing wave [9] pointed the way toward the realization of that goal. Bragg diffraction efficiency can be varied from 0 % to 100 % by simply adjusting the Bragg-pulse duration and/or intensity [10, 11]. This means that Bragg diffraction can function as an ideal, adjustable beam splitter/mirror for a BEC, providing us with the critical atom-optic elements needed for constructing an atom-laser interferometer. In this Letter we report the first demonstration of a Mach-Zehnder interferometer for a BEC using an optical Bragg-diffraction technique. This not only allows novel studies of the coherence properties of a BEC itself, such as vorticity [12], but also paves the way toward extremely sensitive atom-wave interferometers that could be used to detect gravitational waves.

Atom-wave interferometers comprised of optical standing waves can be classified into two types, those based on normal diffraction and those based on Bragg diffraction. In the case of normal diffraction there are many spurious momentum space paths and only a small fraction of atoms (~10%) can contribute to the signal [13]. With Bragg diffraction however, the atomic wave can be coherently split into only two paths, and then coherently recombined. This results in high efficiency [14] where, in principle, all of the atoms can contribute to the signal. Normal and Bragg diffraction based atom-wave interferometers were first demonstrated in 1995 using collimated metastable atomic beams, and fringe contrasts of 10% (normal) and 62% (Bragg) were obtained [13,14]. This is to be compared to our BEC Bragg-diffraction interferometer where all the atoms contribute to the signal due to the condensate's extremely narrow velocity spread. We achieve nearly the maximum contrast of 100 % with a good single-shot visibility.

Bragg diffraction occurs when the atomic matter wave (a BEC nearly at rest) is coherently scattered by a moving, optical, standing wave formed by counter-propagating laser beams of slightly different frequencies. The mechanism of Bragg diffraction can be most easily understood as a two-photon stimulated Raman process [9], where photons from one laser beam are coherently scattered into the other, changing the atomic momentum in the process. The frequency difference between the two counter-propagating Bragg beams is chosen to correspond to the energy difference between the two momentum states. In principle, the initial momentum state $|p = 0\hbar k\rangle$ can be coupled to any momentum state $|p = 2n\hbar k\rangle$ ($n$ integer) using $n^{th}$-order Bragg diffraction. In our case we choose to couple momentum states $|p = 0\rangle$ and $|p = 2\hbar k\rangle$ ($k = 2\pi/\lambda$, and $\lambda = 780$ nm is the laser wavelength). The momentum-space wavefunction of a condensate (initially in state $|p = 0\rangle$) continuously irradiated with Bragg diffraction beams will oscillate between the two coupled momentum states just as if it were a two-level system. The effective oscillation frequency is $\Omega_{eff} = \Omega_1\Omega_2/2\Delta$, where $\Omega_1$ and $\Omega_2$ are the resonant Rabi frequencies of two Bragg beams, and $\Delta$ is the detuning of the beams from the optical transition [11, 15]. As mentioned earlier, an arbitrary percentage of the atoms can be transferred to the $|p = 2\hbar k\rangle$ momentum state by properly adjusting the intensity, detuning, and/or duration of the Bragg pulse.

This closed two-level system can be most easily thought of as a fictitious spin 1/2 system, where we define $|g\rangle \equiv |p = 0\hbar k\rangle$, and $|e\rangle \equiv |p = 2\hbar k\rangle$. As usual, a $\pi/2$ ($\pi$) pulse is one that results in transferring half (all) of the atoms from $|g\rangle$ to $|e\rangle$. More precisely, the state vector of the system under a $\pi/2$-pulse obeys the transformation equations $|g\rangle \rightarrow (|g\rangle - e^{-i\varphi}|e\rangle)/\sqrt{2}$, and $|e\rangle \rightarrow (e^{+i\varphi}|g\rangle + |e\rangle)/\sqrt{2}$. Here $\varphi$ is the phase of the moving standing wave in the center of the initial atomic wavepacket $|g\rangle$ in the middle of the Bragg pulse. Changes in $\varphi$ are measured with respect to the phase of the resonant, moving standing wave.

Successive application of this transformation shows that under a π-pulse $|g\rangle \rightarrow -e^{-i\varphi}|e\rangle$, and $|e\rangle \rightarrow e^{+i\varphi}|g\rangle$. We can therefore use a π/2 (π) pulse as an ideal beamsplitter (mirror) for the condensate.

We prepare a BEC of rubidium atoms using a dc magnetic trap and a standard evaporation strategy [1, 16]. Briefly, we first trap about $10^9$ $^{87}$Rb atoms in an ultra-high vacuum glass cell using a double magneto-optical trap [17]. The atoms are then transferred into a cloverleaf magnetic trap [18] and cooled using rf-induced evaporation. Every 5 minutes we create a BEC containing about $10^5$ atoms in the $5S_{1/2}$ $F = 1$, $m_F = -1$ state, in a trap with radial gradient, axial curvature, and bias field of 1.75 T/m, 185 T/m$^2$, and $10^{-4}$ T, respectively. Due to the geometry of our magnetic trap, the condensate is cigar-shaped with the symmetry axis perpendicular to the direction of gravity. We release the condensate by suddenly switching off the magnetic trap within 200 μs, wait 5 ms for the mean-field driven explosive expansion to subside, and apply a Bragg interferometer pulse sequence. After 20 ms of free evolution to allow the $|g\rangle$ and $|e\rangle$ components enough time to separate significantly in space, an absorption image [1] of the resulting condensate is taken. This yields very clean and unambiguous signals since the action of the interferometer is mapped into the probability of observing atoms in two spatially separated regions.

The moving, optical, standing wave is generated using two counter-propagating laser beams with parallel linear polarization but slightly different frequencies. These two laser beams are derived from a single diode laser using acousto-optic modulators, and the spatial mode of each beam is purified by passing them through single-mode fibers. The light propagates parallel to the symmetry (long) axis of the trapped condensate, and the frequency difference between the two laser beams is $\delta$ = 15 KHz. This relative detuning corresponds to the two-photon recoil energy for rubidium mentioned earlier. Our laser intensities were chosen such that $\Omega_1/2\pi = \Omega_2/2\pi \cong 5$ MHz. To suppress spontaneous emission we used a detuning $\Delta/2\pi = 2$GHz. These parameters result in an effective two-photon Rabi frequency of $\Omega_{eff}/2\pi \cong 6.3$ kHz.

The experimental schematic of our atom interferometer is shown in Fig. 1. The momentum-space condensate wavefunction Ψ, initially in state $|g\rangle$ ($|p = 0 \hbar k\rangle$), is coherently split by the first (π/2) pulse and becomes $\Psi_1= (|g\rangle-|e\rangle)/\sqrt{2}$. Here, without the loss of generality, we have chosen $\varphi = 0$ (when $\varphi \neq 0$ there is an additional, physically meaningless global phase multiplying the final wavefunction). In real space these two different momentum states begin to separate and, after a delay $\Delta T$ (measured from the center of successive Bragg pulses), the second ($\pi$) pulse is applied, producing $\Psi_2= -(|g\rangle+|e\rangle)/\sqrt{2}$. Now the two spatially separated parts of $\Psi_2$ begin to converge because the action of the $\pi$-pulse was effectively that of a mirror in momentum space. After another equal delay $\Delta T$, the two parts of $\Psi_2$ from the two different coordinate-space paths overlap completely, and the third ($\pi/2$) pulse is applied yielding $\Psi_3= -|g\rangle$. In this case the probability of finding the atoms in the initial state $|g\rangle$ after the sequence of three Bragg pulses is $P_g=|\langle\Psi_3|g\rangle|^2=1$, regardless of the delay $\Delta T$ between the pulses. If, however, the phase of the moving, standing wave were altered by $\phi$ before applying the third Bragg pulse, $\Psi'_3= -[(1+e^{i\phi})|g\rangle+(1-e^{-i\phi})|e\rangle]/2$, and $P_g=[1+\cos(\phi)]/2$ ($P_e=[1-\cos(\phi)]/2$). Fringes in the final probability of observing the atoms in the initial (or final) momentum state may therefore be mapped out by varying $\phi$ just before applying the final $\pi/2$ Bragg pulse.

The Bragg-pulse duration (for fixed $\Omega_{eff}$ and $\Delta$) needed to produce a $\pi/2$ ($\pi$) pulse was first measured experimentally. Figures 2a-c show absorption images of the condensate after applying a single Bragg pulse of duration $\tau = 0$, 40 ($\pi/2$-pulse), and 80 ($\pi$–pulse) $\mu$s respectively. The condensate images are elliptical because of its initial anisotropic shape [1,16]. The experimentally determined durations are in excellent agreement with the calculated values of $\tau = 40$ $\mu$s and 80 $\mu$s based on the measured laser intensities and detuning. Once the $\pi/2$ ($\pi$) pulse durations were empirically determined, we proceeded with the actual interferometer experiment (Fig. 1) by applying the $\pi/2-\pi-\pi/2$ sequence of three Bragg pulses. The relative phase $\phi$ of the third $\pi/2$ pulse was experimentally adjusted in the range $0 \leq \phi \leq 3.5\pi$ by the changing the phase of one of the two Bragg beams comprising the moving standing wave with an electro-optic modulator (EOM).

Figure 3 shows the resultant oscillation of the population in the |e> state as a function of ϕ when ΔT = 190 μs (each point represents a single measurement). Note that we observe an interference pattern with almost 100 % contrast. Because the Bragg beams were not perfectly perpendicular to the direction of gravity, the falling condensates see a differential phase shift even when ϕ=0. This results in the small ~0.1π phase shift in the fringe pattern of Fig. 3. This phase shift corresponds to an angle misalignment of a few degrees from perpendicular, which is well within our experimental uncertainty.

The chosen pulse interval of 190 μs corresponds to a 2.2 μm separation of the two arms of the interferometer, or about six times the period of the optical standing wave made by the Bragg beams. Since the size of the condensate in the direction of momentum transfer is roughly 50 μm, ΔT > 5 ms is necessary to separate completely the two arms of the interferometer. However, we found that when ΔT > 2 ms, reproducibility of the interference pattern deteriorates significantly. When ΔT=3 ms, the resulting fringe pattern is completely random, ranging from 0 % to 100 %. This indicates that the recombining wavepackets are fully coherent, even though their relative phase is not well controlled. The phase stability of the Bragg beams was measured with a homodyne detection technique and was found to be shorter than 1 ms. This explains the lack of reproducibility at long ΔT and we suspect that the random relative phase is due to mechanical vibration of the numerous mirrors used for the Bragg beams. We believe that a stable signal contrast of 100 % visibility can be achieved at long ΔT by actively stabilizing these critical Bragg mirrors. This would mean that we could realize a stable interferometer whose two arms were completely separated in space, resulting in a very large enclosed area.

The Bragg interferometer presented here is a novel tool that can be used to study the phase properties of condensates. For example, it should be possible to detect the phase signature of a vortex in a Bose-Einstein condensate with this interferometer [12]. A vortex state could be created in one arm of the interferometer and analyzed by interfering it with the other arm. The resultant momentum-space map of the condensate phase after the final π/2-pulse could then be analyzed with conventional imaging techniques [1]. In addition to studying fundamental properties of condensates, such a BEC interferometer might prove useful in the detection of gravitional waves because of its high sensitivity and high signal-to-noise ratio. Although promising, the mean-field interaction of the atoms themselves may cause interferometer instability at long ΔT because of shot-to-shot ($\sqrt{N}$) variations in the number of atoms in each arm of the interferometer after the first π/2-pulse. We are currently working on stabilizing and measuring the limitations of our Mach-Zehnder atom interferometer.

We thank T. Sugiura for technical assistance and helpful discussions. This work has been supported by a Grant-in-Aid from the Ministry of Education, Science, and Culture, as well as by Core Research for Evolutionary Science and Technology (CREST) of the Japan Science and Technology Corporation (JST).

*Current address: Department of Physics, Gakushuin University, Mejiro 1-5-1, Toshima-ku, Tokyo171-8588, Japan

9

**Figure captions**

**Fig. 1)** Experimental schematic of the π/2–π–π/2 Mach-Zehnder Bragg interferometer.

**Fig. 2)** Absorption images of the Bose condensate taken 20 ms after one Bragg pulse is applied (upper), and density profiles taken through the center of the condensates (lower), as a function of the pulse duration τ.

**Fig. 3)** Population oscillation of the condensate in the $|p = 2\hbar k\rangle$ ($|e\rangle$) momentum state as a function of the phase shift ϕ of the third Bragg pulse. The time interval between the center of successive Bragg pulses is $\Delta T = 190$ μs.

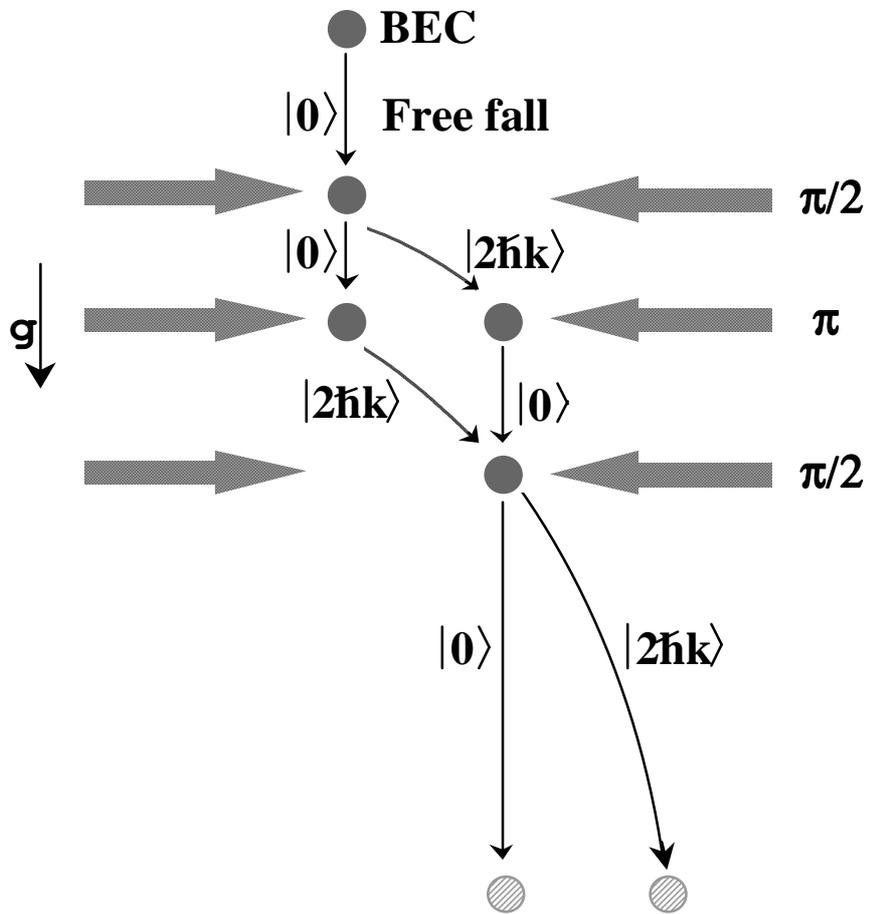

Fig.1 Y. Torii et al. "Mach-Zehnder Bragg interferometer..."

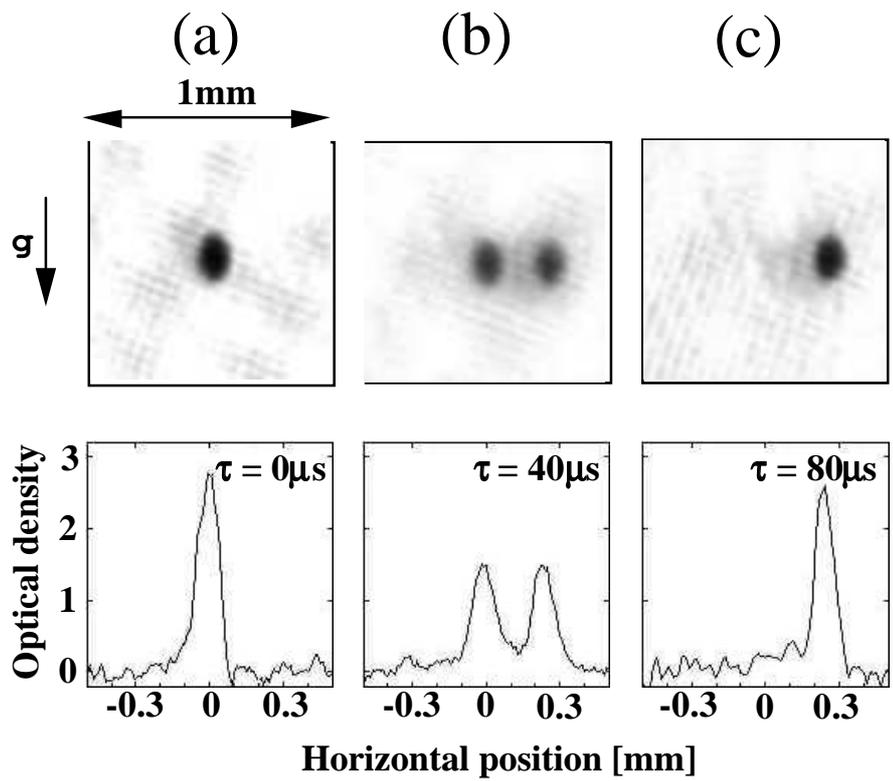

Fig.2 Y. Torii et al. "Mach-Zehnder Bragg interferometer..."

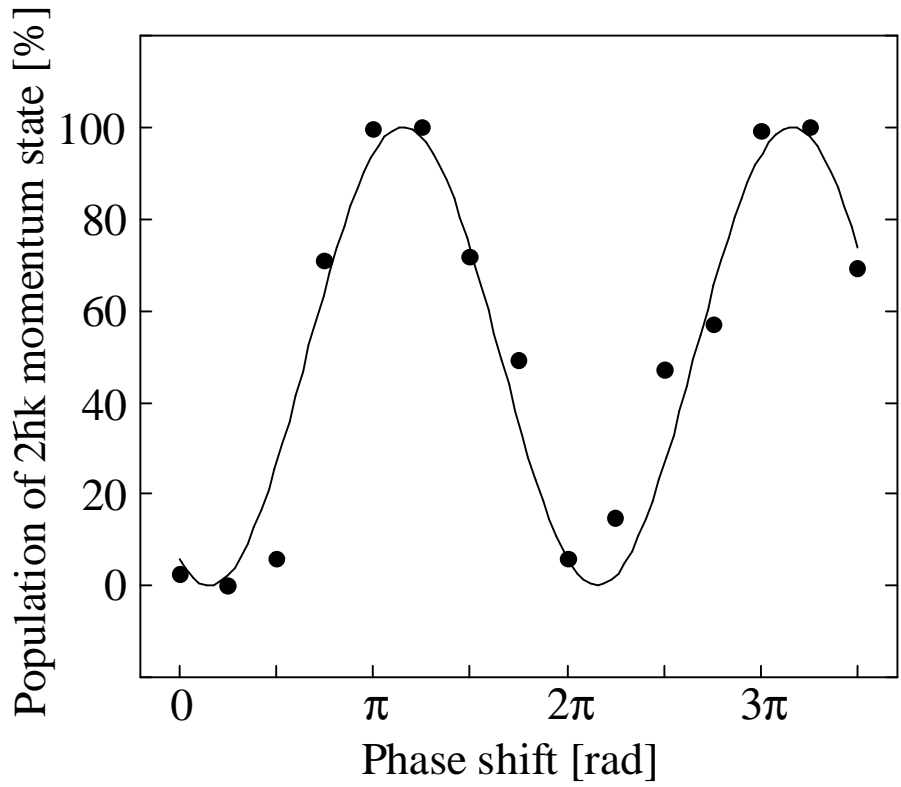

Fig.3 Y. Torii et al. "Mach-Zehnder Bragg interferometer..."